\documentclass[12pt]{article}
\pdfoutput=1

\usepackage{epsf}
\usepackage{amsfonts}

\newcommand\ZZZ{{\hbox{ Z\kern-1.6mm Z}}}
\newcommand{\Iop}{\relax{\rm I\kern-.18em I}}
\newcommand{\Lop}{\relax{\rm I\kern-.18em L}}
\newcommand{\dop}{\relax{\rm I\kern-.8em d}}
\newcommand{\one}{{\hbox{ 1\kern-1.2mm l}}}

\newcommand{\beq}{\begin{equation}}
\newcommand{\eeq}{\end{equation}}
\newcommand{\bea}{\begin{eqnarray}}
\newcommand{\eea}{\end{eqnarray}}
\newcommand{\ra}{\rangle}
\newcommand{\la}{\langle}
\newcommand{\lt}{\left}
\newcommand{\rt}{\right}

\newcommand{\del}{\partial}

\newcommand{\al}{\alpha}

\newcommand{\dlt}{\delta}
\newcommand{\eps}{\epsilon}

\newcommand{\Dlt}{\Delta}

\newcommand{\cE}{{\cal E}}
\newcommand{\cF}{{\cal F}}

\newcommand{\cL}{{\cal L}}

\newcommand{\hR}{\hbox{R}}

\newcommand{\hi}{\hbox{i}}

\newcommand{\mE}{\mathbb{E}}

\newcommand{\mW}{\mathbb{W}}

\newcommand{\rv}{{\hbox{\bf r}}}
\newcommand{\hbd}{{\hbox{\bf d}}}

\begin{document}

{}~
{}~
\hfill\vbox{\hbox{IMSc/2012/3/5}}
\break

\vskip 2cm

\centerline{\Large \bf All order covariant tubular expansion}

\medskip

\vspace*{4.0ex}

\centerline{\large \rm Partha Mukhopadhyay }

\vspace*{4.0ex}

\centerline{\large \it The Institute of Mathematical Sciences}
\centerline{\large \it C.I.T. Campus, Taramani}
\centerline{\large \it Chennai 600113, India}

\medskip

\centerline{E-mail: parthamu@imsc.res.in}

\vspace*{5.0ex}

\centerline{\bf Abstract}
\bigskip

We consider tubular neighborhood of an arbitrary submanifold embedded in a (pseudo-)Riemannian manifold. This can be described by Fermi normal coordinates (FNC) satisfying certain conditions as described by Florides and Synge in \cite{FS}. By generalizing the work of Muller {\it et al} in \cite{muller} on Riemann normal coordinate expansion, we derive all order FNC expansion of vielbein in this neighborhood with closed form expressions for the curvature expansion coefficients. Our result is shown to be consistent with certain integral theorem for the metric proved in \cite{FS}.

\newpage

\tableofcontents

\baselineskip=18pt

\section{Motivation and summary}
\label{s:intro}

As it is well known, Riemann normal coordinate (RNC) system is a very useful tool in differential geometry and general relativity \cite{bishop}-\cite{wald}. In particular, it is used for computing covariant Taylor expansion of tensors around a point in a manifold \cite{QFT1}-\cite{sigma4}.

Fermi normal coordinate (FNC) system is a generalization of RNC in the sense that the RNC-origin, which may be viewed as a zero-dimensional submanifold of the ambient space, is replaced by a higher dimensional one\footnote{FNC was first introduced by E. Fermi for a curve in \cite{fermi}. It was then followed by various other generalizations (see \cite{fermi-gen1}-\cite{fermi-gen3} and references therein). Our consideration, which is same as that of \cite{FS}, will be described in detail in \S \ref{s:FS}.}. It is related to a general coordinate system through the exponential map along the normal directions. The region surrounding the submanifold where FNC is well defined, i.e. where the exponential map is a diffeomorphism, is called a {\it tubular neighborhood}. The existence theorem for such a neighborhood suffices as a powerful tool for various analysis in differential geometry \cite{spivak, bredon}. 

The use of FNC can be found in various areas of physics. Some of them, which we will briefly discuss below, are general relativity, multi-particle and string dynamics in curved spacetime and {\it constrained quantum mechanics}. In all such applications one needs to know explicitly the covariant Taylor expansion coefficients of geometric quantities around the submanifold. The same is true for certain applications outside physics also. For example, see \cite{tubes} which describes how this is used in computing volume of tubes \cite{tube-vol1}-\cite{loader}.

In general relativity one is mainly interested in studying various physical effects around a particle worldline (one dimensional submanifold) in a given background \cite{li1}-\cite{collas}.
In one line of research (\cite{parker1}-\cite{caicedo}) FNC expansion around a worldline has been used to compute curvature corrections to the energy spectrum of hydrogen-like atoms in curved backgrounds. The reason for considering a worldline is the following approximation (see also \cite{collas}): the center of mass (CM) of the system falls freely along a geodesic during the time interval relevant for the study of the system (e.g. an atomic transition). 

The above assumption is physically well-motivated and should work fine when the CM-mass is large. However, it is also a simplifying assumption. Going beyond this assumption requires a drastic change in the analysis where a different tubular geometry around a higher dimensional submanifold becomes relevant. Let us explain this slightly more elaborately. Physically one should expect that the internal dynamics of the atom (or any other composite object) backreacts to the motion of the CM. In other words, it should be possible to derive the geodesic equation for CM at the leading order in a semi-classical expansion\footnote{What we have in mind is an expansion in inverse of the CM-mass. The quantum mechanical $\hbar$ may be set to $1$.} where the CM mass is large. Notice that this is a relativistic bound state problem in curved spacetime. A completely satisfactory description of such a problem has not yet been understood even in flat spacetime \cite{relativistic-bound1, relativistic-bound2}. However, one may expect that the basic geometric structure underlying any manifestly covariant description of an $n$-particle bound state problem in $M$ is the tubular geometry around $\Dlt \hookrightarrow M^n$. Here $M^n := M\times M \times \cdots (n-\hbox{times})$ is the $n$-particle (extended) configuration space and $\Dlt (\cong M)$ is the diagonal submanifold of all possible locations of the CM (assuming equal mass for all the particles). The reason why $\Dlt$ is the subspace where the CM lies has been explained in \cite{semi-classical}.

Recently the above approach has been pursued in an analogous system in \cite{semi-classical}. Here the analogue of a multi-particle bound configuration is a closed string in $M$, the configuration space is the corresponding loop space $LM$, the relativistic quantum theory is the loop space quantum mechanics \cite{semi-classical}-\cite{dwv3} and the analogue of $(\hbox{CM-mass})^{-1}$ is the parameter $\alpha'$. It was shown that in a semi-classical limit ($\alpha' \to 0$) the string wavefunction localizes on $\Dlt (\cong M)$ - the subspace of string CM, which is same as the space of vanishing loops. Moreover, the semi-classical expansion is obtained from the tubular expansion of geometric quantities around $\Dlt \hookrightarrow LM$.\footnote{Therefore, according to this analogy the corrections to the CM motion of any quantum mechanical bound state in $M$ are analogous to $\alpha'$ corrections in string theory.}

Another application of tubular geometry around a higher dimensional submanifold is found in the context of what has been termed in the literature as constrained quantum systems \cite{dacosta, maraner}. Here one considers a non-relativistic classical system in an ambient space with a potential that tries to confine the motion into a submanifold. The idea is to realize this constraint at the quantum mechanical level through localization of wavefunction \cite{mitchell}-\cite{wachsmuth2}. The mechanism of finding the effective theory on the submanifold is very similar to that of the string case discussed above.

As mentioned earlier, the basic raw data that goes into all the above computations are the tubular expansion coefficients of geometric quantities around the submanifold. So far only a few such coefficients have been known. The goal of this paper is to obtain closed form expression for the expansion coefficients of vielbein to all orders in a completely generic situation where an arbitrary submanifold is embedded in a higher dimensional space.

We now briefly discuss the technical points relevant to our analysis. In \cite{FS}, Florides and Synge (FS) constructed the special coordinate system under consideration for an arbitrary submanifold embedded in a higher dimensional (pseudo-)Riemannian space\footnote{The authors of \cite{FS} called it a {\it submanifold based coordinate system}. However, following the modern nomenclature (see, for example \cite{tubes}) we will continue to call it FNC.}. They wrote down the special coordinate conditions in terms of the metric and proved an integral theorem describing its behavior away from the submanifold. We will review the basic results in \S \ref{s:FS}.

In \cite{muller}, Muller, Schubert and van de Ven considered the RNC coordinate conditions written in terms of the vielbein and spin connection. Using certain differential geometric techniques it was possible to write down an integral equation for the vielbein in terms of the Riemann curvature tensor\footnote{As mentioned in \cite{muller}, such coordinate conditions are the gravity analogue of the Fock-Schwinger gauge \cite{schwinger} in gauge theory which can also be used to write down an integral equation for the gauge potential in terms of the field strength \cite{shifman}.}. The authors were able to solve this equation to produce the complete RNC expansion of vielbein with a closed form covariant expression for the curvature expansion coefficients.

Here we will use the same techniques to generalize the results of \cite{muller} to the case of FNC. In particular, we derive the integral equation for the vielbein and closed form covariant expressions for its curvature expansion coefficients in the set up considered in \cite{FS}. The results are different when the vector index of the vielbein takes values along the directions tangential and transverse to the submanifold. The transverse results are exactly the same as that of \cite{muller}, as expected. This will be discussed in \S \ref{s:arbit}. In \S \ref{s:rederive}, we show how our result is consistent with the metric integral theorem of \cite{FS}. We discuss a demonstrative example in \S \ref{s:example} and finally conclude in \S \ref{s:conclusion}.
Explicit numerical results for the expansion of vielbein and metric up to $10$-th order in FNC have been presented in one of the appendices.

\section{Metric-integral-theorem due to Florides and Synge}
\label{s:FS}

We consider an arbitrary $D$-dimensional submanifold $M$ embedded in a higher dimensional pseudo-Riemannian space $L$ of dimension $d$.  Our notation for indices is as follows: Greek indices ($\alpha, \beta, \cdots$) run over $D$ dimensions, capital Latin indices ($A, B, \cdots$) run over $(d-D)$ transverse dimensions and small Latin indices ($a, b, \cdots$), over all dimensions. Other notations and conventions essential for the rest of our discussion are presented in appendix \ref{a:notations}. 

The work of \cite{FS} proved the existence of a coordinate system $z^a = (x^{\alpha}, y^A)$ (that will be called FNC following \cite{tubes}), where $x^{\alpha}$ is a general coordinate system in $M$ whose embedding is given by,
\bea
y^A = 0~,
\label{submanifold-eq}
\eea
and the following conditions are satisfied for the transverse coordinates,
\bea
g_{a B}(x, y)y^B &=& \eta_{a B} y^B~, 
\label{coord1}
\eea
where $g_{ab}(x,y)$ is the metric tensor in FNC and $\eta_{ab}$ is as define in eq.(\ref{eta}). 

The following integral theorem was proved in \cite{FS},
\bea
g_{AB}(x, y) &=& \eta_{AB} + 2 y^C y^D \int_0^1 dt~ F_1(t) l_{ACDB}(x, ty)~, \cr
g_{A\beta}(x, y) &=& y^C \underline{g_{A\beta, C}} + 2 y^C y^D \int_0^1 dt~ F_2(t) l_{ACD\beta}(x, ty)~, \cr
g_{\alpha \beta} (x, y) &=& G_{\alpha \beta}(x) + y^C \underline{g_{\alpha \beta,
  C}} + 2 y^C y^D \int_0^1 dt~ F_3(t) l_{\alpha CD\beta}(x, ty)~.
\label{int-g}
\eea
For any function $f(x,y)$ in $L$, we have defined: $\underline{f} \equiv f(x,0)$. A comma in the suffix indicates ordinary derivative with respect to the argument. For example,
\bea
\underline{g_{a \beta, C}} = \lim_{y\to 0} {\del \over \del y^C}g_{a \beta}(x, y)~.
\eea
$G_{\alpha \beta} = \underline{g_{\alpha \beta}}$ is the induced metric on $M$ and the functions $F_i(t)$ ($i=1,2,3$) are defined as follows,
\bea
F_1(t) = t(1-t)~, \quad F_2(t) = {1\over 2}(1-t^2)~, \quad F_3(t) = 1-t~.
\label{F-def}
\eea
Finally,
\bea
l_{acdb}(x,y) = {1\over 2} (g_{ab,cd} + g_{cd, ab} - g_{ad,cb} - g_{cb, ad}) (x, y)~,
\label{l-def}
\eea
is the linear part of the covariant Riemann curvature tensor.

\section{FNC expansion in arbitrary tubular neighbourhood}
\label{s:arbit}

The above metric-integral-theorem involves the linear part of the curvature tensor. This form is not very useful for deriving the curvature expansion. Below we will use the technique of \cite{muller} to derive the curvature expansion for vielbein.
In \S \ref{s:rederive} we will show how our result is consistent with that of FS.

\subsection{Integral equations for vielbein}
\label{ss:int}

We will first derive the integral equations satisfied by the vielbein (eqs.(\ref{int-viel}) below). To this end, we define, following \cite{muller}, the {\it radial vector field} (see appendix \ref{a:notations} for our notations),
\bea
{\hbox{\bf r}} = y^A \cE_A~.
\eea
Then the coordinate condition (\ref{coord1}) can be rewritten in terms
of the vielbein and the connection one-form in the following way,
\bea
\hi_{\rv} \hat \cE^{(a)} &=& \dlt^{a}_B y^B~, 
\label{coord-viel} \\
\hi_{\rv} \omega^{(a)}{}_{(b)} &=& 0~,
\label{coord-spin}
\eea
where $\hi_{\rv}$ denotes the interior product \cite{nakahara} with respect to the vector field $\rv$ and $\dlt^a{}_b$ is the Kronecker delta function\footnote{To see how (\ref{coord-viel}, \ref{coord-spin}) and
(\ref{coord1}) are equivalent one first shows that (\ref{coord1})
implies,
\bea
\gamma^a_{BC}(x, y) y^B y^C = 0~,
\label{gamma-cond}
\eea
and {\it vice versa} ($\gamma^a_{bc}$ being the Christoffel symbols as defined in eq.(\ref{gamma}) ). One then rewrites the above equations in terms of the vielbein and spin connection coefficients to derive (\ref{coord-spin}) and,
\bea
\del_A e^{(a)}{}_B(x,y) y^A y^B = 0~,
\eea
which, in turn, is solved by (\ref{coord-viel}).}.

The next step is to relate the second order Lie derivative of $\hat \cE^{(a)}$ and the curvature two-form. Using eqs. (\ref{structure-theta}, \ref{coord-spin}) one writes,
\bea
\cL_{\rv} \hat \cE^{(a)} = \omega^{(a)}{}_{(b)} \hi_{\rv} \hat \cE^{(b)} + d \hi_{\rv} 
\hat \cE^{(a)}~.
\eea
Then using the following result, which can be easily proved by using (\ref{coord-viel}),
\bea
\cL_{\rv} d \hi_{\rv} \hat \cE^{(a)} = d \hi_{\rv} \hat \cE^{(a)}~,
\eea
one arrives at,
\bea
\cL_{\rv}(\cL_{\rv}-1) \hat \cE^{(a)} &=& \cL_{\rv} \omega^{(a)}{}_{(b)} \hi_{\rv} \hat \cE^{(b)}~.
\label{lie2theta}
\eea
Finally, one calculates two sides of the above equation independently. The left hand side is directly calculated by noting,
\bea
\hat \cE^{(a)} = e^{(a)}{}_{\alpha}(x,y) \cE^{\alpha} + e^{(a)}{}_A(x,y) \cE^A~,
\eea
and the right hand side can be calculated by using eqs.(\ref{structure-omega}, \ref{coord-spin}). This leads to the following second order differential equation for the vielbein,
\bea
\hbd (\hbd + \mathop{\eps}^b) e^{(a)}{}_b(x,y) = \rho^{(a)}{}_{(c)}(x,y;y) e^{(c)}{}_b(x,y)~,
\label{diff-viel}
\eea
where 
\bea
\rho^{(a)}{}_{(b)} (x,y; \tilde y) \equiv r^{(a)}{}_{CD (b)}(x,y) \tilde y^C \tilde y^D~.
\eea
We convert the non-coordinate to coordinate indices and {\it vice-versa} by using the vielbein and its inverse. For example,
\bea
r^{(a)}{}_{CD(b)} &=& r^{(a)}{}_{(c)(d)(b)} e^{(c)}{}_C e^{(d)}{}_D~.
\eea
Moreover, for any function $f(x,y)$ we have defined,
\bea
\hbd f(x,y) = y^A \del_A f(x,y)~,
\label{db}
\eea
and,
\bea
\mathop{\eps}^b = \lt\{\begin{array}{rl}
1 & \quad \hbox{when } b=B ~, \cr
-1 & \quad \hbox{when } b=\beta~.
\end{array} \rt.
\label{eps}
\eea

Following \cite{muller}, we Taylor expand both sides of (\ref{diff-viel}) around $y=0$
and equate the $n$-th order terms. This gives,
\bea
\hbd_0^n e^{(a)}{}_B(x, 0) &=& {n-1\over n+1} \hbd_0^{n-2}
\lt[\rho^{(a)}{}_{(c)}(x,0;y) e^{(c)}{}_B(x, 0) \rt]~, \quad \quad n
\geq 1~, \cr && \cr
\hbd_0^n e^{(a)}{}_{\beta}(x, 0) &=& \hbd_0^{n-2} \lt[\rho^{(a)}{}_{(c)}(x,0;y)
e^{(c)}{}_{\beta}(x, 0)\rt]~,  \quad \quad n\geq 2~,
\label{d0^n-e-ab}
\eea
where we have introduced a new notation,
\bea
\hbd_0^n [f(x,0; y) g(x,0)] &=& y^{A_1} \cdots y^{A_n} 
\lim_{\tilde y \to 0} {\del \over \del \tilde y^{A_1}} \cdots {\del \over \del
  \tilde y^{A_n}} [f(x,\tilde y; y) g(x, \tilde y)]~.
\eea

The integral forms that solve the above equations are as follows,
\bea
e^{(a)}{}_B(x,y) &=& \dlt^{a}_B + \int_0^1 dt~ F_1(t)
\rho^{(a)}{}_{(c)}(x, ty;y) e^{(c)}{}_B(x, ty)~, \cr
e^{(a)}{}_{\beta}(x,y) &=& \underline{e^{(a)}{}_{\beta}} + y^A \underline{ e^{(a)}{}_{\beta, A}} +
\int_0^1 dt~ F_3(t) \rho^{(a)}{}_{(c)}(x, ty;y) e^{(c)}{}_{\beta}(x, ty)~. \cr &&
\label{int-viel}
\eea
The first term in the first equation and the first two terms in the second equation are ``initial conditions". The former is obtained by contracting the first equation with $y^B$, using (\ref{coord-viel}) and then noticing that the integral term does not contribute under such a contraction because of the anti-symmetry property of the Riemann tensor. For the latter we have\footnote{In the rest of the discussion in this subsection, we will refrain from explicitly writing down the arguments. It is understood that a geometric quantity denoted by a lower case symbol without a bar has an argument $(x, y)$ and the same with a bar has an argument $(x)$.},
\bea
\underline{e^{(\alpha)}{}_{\beta}} = E^{(\alpha)}{}_{\beta}(x) ~, \quad \underline{e^{(A)}{}_{\beta}}= 0~, \quad y^A \underline{e^{(a)}{}_{\beta, A}} = y^A \underline{\omega_{\beta}{}^{(a)}{}_A}~,
\label{in-cond2}
\eea
where $E^{(\alpha)}{}_{\beta}$ is the vielbein of the induced metric $G_{\alpha \beta}$ on the submanifold. While the first equation is obvious, the second equation is required by consistency with (\ref{coord1}). The third equation is obtained as follows. Using the following relation\footnote{The LHS of eq.(\ref{tetrad-postulate}) is usually called the {\it total covariant derivative} of vielbein. The vanishing of this is sometimes referred to as {\it vielbein postulate}. However, this equation directly follows from (\ref{spin-connection}) and therefore can be viewed as the defining equation for spin connection. } ,
\bea
\del_a e^{(b)}{}_c - \gamma^d_{ac} e^{(b)}{}_d + \omega_a{}^{(b)}{}_{(e)} e^{(e)}{}_c = 0~,
\label{tetrad-postulate}
\eea
and (\ref{coord-viel}) one gets on the submanifold,
\bea
\underline{\gamma^d_{\beta A} e^{(a)}{}_d} = \underline{\gamma^d_{A \beta} e^{(a)}{}_d }
= \underline{\omega_{\beta}{}^{(a)}{}_A} ~,
\label{gamma-omega}
\eea
Then using (\ref{tetrad-postulate}), (\ref{gamma-omega}) and (\ref{coord-spin}) one arrives at 
the third equation in (\ref{in-cond2}).

\subsection{Closed form expressions for curvature expansion-coefficients}
\label{ss:coeff}

We write,
\bea
e^{(a)}{}_b(x, y) &=& \sum_{p=0}^{\infty} e_p^{(a)}{}_b(x,y)~,
\label{e_p-def}
\eea
where $e_p$ is the contribution at $p$-th order in curvature, such that,
\bea
e_0^{(a)}{}_B &=& \dlt^{a}_B~, \cr
e_0^{(a)}{}_{\beta} &=& \lt\{ \begin{array}{ll}
E^{(\alpha)}{}_{\beta} + \underline{ \omega_{\beta}{}^{(\alpha)}{}_C} y^C  ~, & \hbox{ for } a=\alpha ~, \\ & \\
\underline{ \omega_{\beta}{}^{(A)}{}_C} y^C ~, & \hbox{ for } a = A ~.
\end{array} \rt.
\label{e_0}
\eea
To obtain the curvature-expansion we proceed as follows. First we use (\ref{e_p-def}) in eqs.(\ref{int-viel}) iteratively to obtain, for $p \geq 1$, 
\bea
e_p^{(a)}{}_b (x, y) &=& \int_0^1 dt_1 (t_1^{2p-2} F_q(t_1) ) \int_0^1 dt_2 (t_2^{2p-4} F_q(t_2) ) \cdots \int_0^1 dt_p  (t_p^0 F_q(t_p))   \cr
&&
[ \rho(x, t_1 y, y) \rho(x, t_1 t_2 y ; y) \cdots 
\rho(x, t_1 \cdots t_p y ; y) e_0(x, t_1\cdots t_p y) ]^{(a)}{}_b ~,
\eea
where $q = 1, 3$ for $b=B,  \beta$ respectively. Then we Taylor expand each $\rho$ factor (around origin of the second argument) in the above equations to get,
\bea
e_p^{(a)}{}_B(x,y) &=& \sum_{s_1, \cdots , s_p \geq 0 } \cF^{(p)}_1(s_1,s_2, \cdots ,s_p) \lt[(y.\nabla)^{s_1} \rho(x,0;y) \cdots (y.\nabla)^{s_p}
\rho(x,0;y) \rt]^{(a)}{}_{(b)} \underline{ e^{(b)}{}_B } ~, \cr
e_p^{(a)}{}_{\beta}(x,y) &=& \sum_{s_1, \cdots , s_p \geq 0 } \cF^{(p)}_3(s_1,s_2, \cdots
,s_p) \lt[(y.\nabla)^{s_1} \rho(x,0;y) \cdots (y.\nabla)^{s_p}
\rho(x,0;y) \rt]^{(a)}{}_{(b)} \underline{e^{(b)}{}_{\beta}}  \cr
&& + \sum_{s_1, \cdots , s_p \geq 0 } \cF^{(p)}_1(s_1,s_2, \cdots,s_p) \lt[(y.\nabla)^{s_1} \rho(x,0;y) \cdots (y.\nabla)^{s_p}
\rho(x,0;y) \rt]^{(a)}{}_{(b)} \underline{\omega_{\beta}{}^{(b)}{}_C } y^C ~, \cr &&
\label{e_p-result}
\eea
where,
\bea
\cF^{(p)}_{q}(s_1,s_2, \cdots ,s_p) &=& {1\over s_1! \cdots s_p!} \int_0^1 dt_1
t_1^{s_1+\cdots +s_p +2p-2} F_{q}(t_1) \int_0^1 dt_2 t_2^{s_2+\cdots +s_p
  +2p-4} F_{q}(t_2) \cr
&& \cdots \int_0^1 dt_p  t_p^{s_p} F_{q}(t_p)~, \quad \quad \quad \quad \quad \quad
q =1, 3~,
\eea
and $F_1(t)$ and $F_3(t)$ are defined in eqs.(\ref{F-def}). Explicit calculations give the following results,
\bea
\cF_1^{(p)}(s_1,s_2, \cdots ,s_p) &=& {C^{(p)}_1(s_1, s_2,\cdots , s_p) \over (s_1+ s_2+\cdots
  +s_p+ 2p+1)!}~, \cr
\cF_3^{(p)}(s_1,s_2, \cdots ,s_p) &=& {C^{(p)}_3(s_1, s_2, \cdots , s_p) \over (s_1+ s_2 +\cdots
  +s_p+ 2p)!}~,
\eea
where,
\bea
C^{(p)}_1(s_1, s_2, \cdots , s_p) &=& C^{s_1+s_2+\cdots +s_p+2p-1}_{s_1} C^{s_2+s_3+\cdots
  +s_p+2p-3}_{s_2} \cdots C^{s_p+1}_{s_p}~, \cr
C^{(p)}_3(s_1, s_2, \cdots , s_p) &=& C^{s_1+ s_2+ \cdots +s_p+2p-2}_{s_1} C^{s_2+ s_3+ \cdots
  +s_p+2p-4}_{s_2} \cdots 1~,
\eea
$C^n_r$ being the binomial coefficients.

In order to present the results in convenient matrix forms we introduce $d\times d$ matrices 
$\underline{ \mE^{\parallel} }$, $\underline{ \mE^{\perp} }$, $\mW$ and $\hbox{R}_{2+s}(x,y)$ such that their elements are given as follows,
\bea
[\underline{ \mE^{\parallel} } ]_{ab} &=& \lt\{
\begin{array}{ll}
E^{(\al)}{}_{\beta}   & \hbox{ for } a = \al, ~ b= \beta \cr
0 & \hbox{ otherwise } ~, 
\end{array}
\rt. 
\eea
\bea
[\underline{ \mE^{\perp} } ]_{ab} &=& \lt\{
\begin{array}{ll}
\dlt^A{}_B   & \hbox{ for } a = A , ~ b= B \cr
0 & \hbox{ otherwise } ~, 
\end{array}
\rt. 
\eea
\bea
[\mW  ]_{ab} &=& \lt\{
\begin{array}{ll}
\underline{ \omega_{\beta}{}^{(a)}{}_C } y^C  & \hbox{ for } b= \beta \cr
0 & \hbox{ otherwise } ~, 
\end{array}
\rt. 
\eea
\bea
[\hbox{R}_{2+s}(x, y)]_{a b} &=& (y.\nabla)^s \rho^{(a)}{}_{(b)}(x,0;y)~.
\eea
Notice that $\hbox{R}_{2+s}$ is linear in curvature, but $(s+2)$-th order in $y$. Using these we further define the following two matrices:
\bea
\mE_{q}(x,y) = \Iop + \sum_{p=1}^{\infty} \sum_{s_1,\cdots, s_p \geq 0} \cF^{(p)}_{q}(s_1,\cdots,s_p) \hR_{2+s_1}(x, y)\cdots \hR_{2+s_p}(x, y)~, \quad q =1, 3~,
\eea
where $\Iop$ is the $d\times d$ identity matrix. The vielbein matrix is given by,
\bea
\mE(x,y) &=& \mE^{\parallel}(x,y) + \mE^{\perp}(x,y)~, \cr
\mE^{\parallel}(x,y) &=& \mE_3(x,y) \underline{\mE^{\parallel}} + \mE_1(x,y) \mW ~,
\quad \mE^{\perp}(x,y) = \mE_1(x,y) \underline{ \mE^{\perp}} ~,
\label{eprl-eprp}
\eea
such that the nonzero elements of $\mE^{\parallel}(x,y)$ and $\mE^{\perp}(x,y)$ are given by $e^{(a)}{}_{\beta}(x,y)$ and $e^{(a)}{}_B(x,y)$ respectively.

The metric is given by,
\bea
g_{ab}(x,y) = e^{(a')}{}_a(x,y) \eta_{(a'b')} e^{(b')}{}_b(x,y)~.
\eea
In matrix form we write,
\bea
g(x,y) = \mE^{\rm T}(x,y) \eta \mE(x,y) = g^{\parallel} + h + h^{\rm T} + g^{\perp}~,
\eea
where
\bea
g^{\parallel} &=& (\mE^{\parallel})^{\rm T} \eta \mE^{\parallel}  
= (\underline{ \mE^{\parallel} })^T X \underline{ \mE^{\parallel} } 
+ (\underline{ \mE^{\parallel} })^T Y \mW + \mW^T Y^T \underline{ \mE^{\parallel} } + \mW^T Z \mW ~, \cr
h &=& (\mE^{\parallel})^{\rm T} \eta \mE^{\perp} 
= (\underline{ \mE^{\parallel} })^T Y \underline{ \mE^{\perp } } + \mW^T Z  \underline{\mE^{\perp} } ~, \cr
g^{\perp} &=& (\mE^{\perp})^{\rm T} \eta \mE^{\perp} = ( \underline{ \mE^{\perp}} )^T Z  \underline{ \mE^{\perp}}  ~,
\label{gprl-h-gprp}
\eea
are all $d\times d$ matrices whose non-zero elements are,
\bea
[g^{\parallel}]_{\alpha \beta} = g_{\alpha \beta}(x,y)~, \quad [h]_{\alpha B} = g_{\alpha B}(x,y) ~, \quad [g^{\perp}]_{AB} = g_{AB}(x,y)~.
\eea
Equations (\ref{eprl-eprp}) and (\ref{gprl-h-gprp}) imply,
\bea
X = (\bar \mE_3)^T \mE_3~, \quad Y =  (\bar \mE_3)^T \mE_1 ~, \quad  Z = (\bar \mE_1)^T \mE_1 ~.
\eea
where,
\bea
{\bar \mE}_{q} = \eta \mE_{q}~.
\eea

We will display explicit results for the matrices $\mE_1$, $\mE_3$, $X$, $Y$ and $Z$ up to $10$-th order in $y$ in appendix \ref{a:coeff}.

\section{Alternative proof of Florides-Synge theorem}
\label{s:rederive}

Here we prove the integral theorem in (\ref{int-g}) starting from our result in (\ref{diff-viel}). Analogous integral equations for vielbein, namely (\ref{int-viel}), were derived from a set of equations (\ref{d0^n-e-ab}) describing all the necessary transverse derivatives of the vielbein evaluated on the submanifold. We will first derive the analogue of eqs.(\ref{d0^n-e-ab}) for the metric (eqs.(\ref{d0^n-g}) below). The integral equations in (\ref{int-g}) will then follow as solutions to eqs.(\ref{d0^n-g}).
  
Writing $g_{ab}=e_a.e_b  = \eta_{(cd)} e^{(c)}_{~~a} e^{(d)}_{~~b}$ and using (\ref{diff-viel}) we get,
\bea
\hbd^2 g_{ab} + \mathop{\eps}^a \hbd e_a.e_b + \mathop{\eps}^b e_a.\hbd
e_b = 2(r_{aCDb}y^C y^D + \hbd e_a.\hbd e_b)~.
\label{d^2g}
\eea
Next we show that the right hand side receives contribution only from the linear part of the curvature tensor, i.e., 
\bea
r_{aCDb}y^C y^D + \hbd e_a.\hbd e_b = l_{aCDb}y^C y^D~,
\label{r-l}
\eea
where $l_{abcd}$ is as defined in eq.(\ref{l-def}). To establish the above equation we first write,
\bea
r_{abcd} = l_{abcd} + q_{abcd}~,
\eea
where the quadratic part of the curvature tensor is given by,
\bea
q_{abcd}  = g_{ef} (\gamma^e_{~ad} \gamma^f_{~bc} - \gamma^e_{~ac}
\gamma^f_{~bd})~.
\label{q-def}
\eea
The condition in (\ref{gamma-cond}) implies that the first term does not contribute in the contraction $q_{aCDb}y^C y^D$. The contribution from the second term can be calculated by using,
\bea
\del_a g_{bC} y^C = (\eta_{bC} -g_{bC}) \dlt^C_a~,
\eea
which follows from eq.(\ref{coord1}). The result is,
\bea
q_{aCDb} y^C y^D &=& -{1\over 4} g^{cd} (\hbd g_{ca} + g_{aC} \dlt^C_{~c} - g_{cC}
\dlt^C_{~a}) (\hbd g_{db} + g_{bD} \dlt^D_{~d} - g_{dD} \dlt^D_{~b}) ~, \cr
&=& -\hbd e_a.\hbd e_b~,
\eea
establishing (\ref{r-l}). To arrive at the second line we used the following result,
\bea
\hbd e_a.e_b - e_a.\hbd e_b &=& y^C(g_{bd} \gamma^d_{~Ca} - g_{ad}
\gamma^d_{~Cb})~, \cr
&=& g_{aC}\dlt^C_{~b} - g_{bC}\dlt^C_{~a} ~,
\label{de.e}
\eea
where the first line is obtained by using an analogue of eq.(\ref{tetrad-postulate}) with indices  properly placed and (\ref{coord-spin}). The second line results from a direct calculation using the explicit form of the Christoffel symbols in terms of metric.

Using (\ref{r-l}) and (\ref{de.e}) in (\ref{d^2g}) we now rewrite the relevant second order differential equations for various metric components in the following way,
\bea
\hbd^2 g_{AB} + \hbd g_{AB} &=& 2 y^C y^D l_{ACDB}~, \cr
\hbd^2 g_{A\beta} - g_{A\beta} &=& 2 y^C y^D l_{ACD\beta}~,\cr
\hbd^2 g_{\alpha \beta} - \hbd g_{\alpha \beta} &=& 2 y^C y^D l_{\alpha CD \beta}~.
\eea
Following the method of \S \ref{ss:int}, we Taylor expand each of the above equations around $y=0$ to find the analogues of eqs.(\ref{d0^n-e-ab}). The results are as follows,
\bea
\hbd_0^n g_{AB}(x, 0) &=& 2 {n-1 \over n+1} y^C y^D \hbd_0^{n-2}
l_{ACDB}(x, 0)~, \quad n \geq 1~, \cr && \cr
\hbd_0^n g_{A\beta}(x, 0) &=& {2n \over n+1} y^C y^D \hbd_0^{n-2}
l_{ACD\beta}(x, 0)~, \quad n\geq 2~, \cr && \cr
\hbd_0^n g_{\alpha \beta}(x, 0) &=& 2 y^C y^D \hbd_0^{n-2} l_{\alpha CD
  \beta}(x, 0)~, \quad n\geq 2~.
\label{d0^n-g}
\eea
The integral equations that solve the above equations are precisely the ones in eqs.(\ref{int-g}).

\section{An example}
\label{s:example}

Here we briefly discuss a demonstrative example of our general result. In \cite{exactFermi} Klein and Collas discussed a class of physically interesting backgrounds. For a specific worldline the exact FNC was constructed and the metric was computed in that system. Here we will demonstrate how our result reproduces the same metric up to second order in Fermi expansion.

In an {\it a priori} coordinate system $\bar z^{\bar a} = (z^0, z^i)$, ($i = 1, \cdots d-1$)\footnote{Our notation is different from \cite{exactFermi}. We leave the total spacetime dimension $d$ arbitrary.}, the metric is given by,
\bea
ds^2 &=& -(1 - f(z^i)) dz^0 dz^0 + \sum_{i=1}^{d-1} dz^i dz^i + [(1-k \bar l^2)^{-1} - 1] d\bar l d \bar l~,
\label{metric-apriori}
\eea
where, $\bar l^2  = \sum_i (z^i)^2$. The local coordinate chart is defined over a region such that $(1-k\bar l^2)>0$. Furthermore, the function $f(z^i)$ satisfies the following conditions, 
\bea
f(0) = 0~, \quad \bar \del_i f(0) = 0~,
\eea
and $(1-f(z^i))$ remains positive within the local chart. The one-dimensional submanifold under consideration is given by,
\bea
z^0 = x ~, \quad z^i = 0~.
\label{worldline}
\eea 

According to the general result of the present paper, the FNC expansion of various components of the metric are given by the following expressions up to second order,
\bea
g_{00} &=& G_{00} + 2 \underline{\omega_{00 C}} y^C  + \underline{(\omega_0{}^a{}_C \omega_{0 a D} + r_{0CD0})} y^C y^D ~, \cr
g_{0 B} &=& \underline{\omega_{0 BC}} y^C + {2\over 3} \underline{r_{0 CDB}} y^C y^D ~, \cr
g_{AB} &=& \eta_{AB} + {1\over 3} \underline{r_{ACDB}} y^C y^D ~,
\label{g-expansion}
\eea
where $G_{00}=-1$ is the induced metric on the worldline and the signature is given by $\eta = diag(-1, 1, \cdots ,1)$. Following our notation in this paper, in the above equations we have used lower case symbols to denote tensors in FNC. To denote tensors in {\it a priori} system we will use the same symbols with a $\bar{}$ at the top.

In order to evaluate the above expansion, one has to relate the expansion coefficients to tensors in {\it a priori} system which can then in turn be evaluated from the metric in (\ref{metric-apriori}) in a straightforward manner. Although the complete transformation relating $\bar z^{\bar a}$ and $z^a$ has been found in \cite{exactFermi}, for our purpose only the Jacobian matrix evaluated at (\ref{worldline}) is needed. It turns out that this is simply given by the identity matrix. Therefore, for example,
\bea
\underline{\omega_{abc}} &=& \dlt^{\bar a}_a \dlt^{\bar b}_b \dlt^{\bar c}_c \underline{\bar{\omega}_{\bar a \bar b \bar c}}~.  
\eea
The specific results that we need are as follows,
\bea
\underline{\bar \omega_{0 \bar a k}} &=& \underline{\bar e_{(\bar b) \bar a} (-\bar \del_0 \bar e^{(\bar b)}{}_k + \bar \gamma^{\bar d}_{0k} \bar e^{(\bar b)}{}_{\bar d} )} = 0~, \cr \underline{\bar r_{0jk0}} &=& {1\over 2} \underline{\bar \del_j \bar \del_k f}~, \quad \underline{\bar r_{ijk0}} = 0~, \quad \underline{\bar r_{ijkl}} = k (\eta_{ik} \eta_{jl} - \eta_{il} \eta_{jk}) ~,
\eea
where in the first line we have used the fact that the metric in (\ref{metric-apriori}) and hence the vielbein is independent of $z^0$ and that $\underline{\bar \gamma^{\bar a}_{0 k}} = 0$ which can easily be verified through explicit computation.

Using the above results we find,
\bea
g_{00} &=& -1 + {1\over 2} \underline{\del_C \del_D f } y^C y^D  ~, \cr 
g_{0 B} &=& 0~, \cr 
g_{AB} &=& \eta_{AB} - {k\over 3} (\eta_{AB} \eta_{CD} - \eta_{AC} \eta_{BD} )y^C y^D ~.
\eea
This is precisely the results in eqs.(15 - 20) in \cite{exactFermi} expanded up to quadratic order in our notation.

\section{Conclusion}
\label{s:conclusion}

The result derived in this work can be viewed as a general theorem. It says that given any tubular neighborhood, it is possible to construct a special coordinate system as described by eqs.(\ref{submanifold-eq}, \ref{coord1}) such that the vielbein components in this system satisfy the integral equations as given in (\ref{int-viel}) and have the Taylor expansions as given in eqs.(\ref{e_p-def}, \ref{e_0}, \ref{e_p-result}). 

As demonstrated in \S \ref{s:example}, this can be used to compute FNC expansion of tensors when the metric and the submanifold under question are known in an arbitrary {\it a priori} system. The above example is particularly simple because of its special nature. It will be useful to have a completely general framework that works for an arbitrary submanifold. 

One specific example is the multi-particle and string dynamics in $M$ as mentioned in \S \ref{s:intro}. Since both $M^n$ and $LM$ are constructed entirely using $M$, all the geometric properties of these two spaces are expressible in terms of the geometric data of $M$ ({\it $M$-data}). This is true, for example, for the tubular expansion of the action or the Hamiltonian that describe a system with the above two spaces as configuration space. The problem at hand is therefore to express such tubular expansions in terms of $M$-data. Investigation to answer such questions is in progress \cite{progress}.

\appendix

\section{Notations and conventions}
\label{a:notations}

We denote the coordinate and non-coordinate bases of the tangent space by $\cE_a$ and $\hat \cE_{(a)}$ respectively,
\bea
\cE_{b} = e^{(a)}{}_b \hat \cE_{(a)} ~,
\eea
$e^{(a)}{}_b$ being the vielbein-components. We denote by $\eta$ the diagonal matrix with indicators of the non-coordinate basis as diagonal elements,
\bea
\eta_{a b} = \la \hat \cE_{(a)}, \hat \cE_{(b)} \ra.
\label{eta}
\eea
The bases dual to $\cE_{a}$ and $\hat \cE_{(a)}$ are denoted by $\cE^a$ and $\hat \cE^{(a)}$ respectively. We consider the torsion-less situation and denote the Christoffel symbols by $\gamma^a_{~bc}$,
\bea
\nabla_b \cE_c \equiv \nabla_{\cE_b} \cE_c = \gamma^{a}_{~bc} \cE_a~,
\label{gamma}
\eea
where $\nabla$ is the Levi-Civita connection. The connection one-form $\omega_{~~(c)}^{(b)} = \omega_{(a)~~(c)}^{~~(b)} \hat \cE^{(a)}$ is given by,
\bea
\nabla_{(a)} \cE_{(c)} \equiv \nabla_{\hat \cE_{(a)}} \cE_{(c)} = \omega_{(a)~~(c)}^{~~(b)} \cE_{(b)}~.
\label{spin-connection}
\eea
Cartan's structure equations in our case are given by,
\bea
d \hat \cE^{(a)} + \omega^{(a)}{}_{(b)}\wedge \hat \cE^{(b)} &=& 0~, 
\label{structure-theta} \\
d \omega^{(a)}{}_{(b)} + \omega^{(a)}_{~~(c)}\wedge \omega^{(c)}_{~~(b)} &=& r^{(a)}_{~~(b)}~,
\label{structure-omega}
\eea
where $r^{(a)}_{~~(b)} = {1\over 2} r^{(a)}_{~(b)(c)(d)}\hat
\theta^{(c)}\wedge \hat \theta^{(d)}$ is the curvature two-form,
\bea
r^{(a)}_{~(b)(c)(d)} = \la \hat \cE^{(a)}, (\nabla_{(c)} \nabla_{(d)} - \nabla_{(d)} \nabla_{(c)} - \nabla_{[\hat \cE_{(c)}, \hat \cE_{(d)}]})\hat \cE_{(b)} \ra ~,
\eea
$[. , .]$ being the Lie bracket \cite{nakahara}.

\section{Coefficients up to tenth order}
\label{a:coeff}

Here we will display the expansion coefficients up to $10$-th order in $y$ for the matrices $\mE_1$, $\mE_3$, $X$, $Y$ and $Z$. We first expand these matrices in the following way,
\bea
\mE_{q} &=& \Iop + \sum_{p=2}^{\infty} \mE_{q}^{(p)} (x, y)~, \quad \quad q = 1, 3~, \cr
X &=& \eta + \sum_{p=2}^{\infty} X^{(p)} = \eta + \sum_{p=2}^{\infty} (\bar{\mE}_3^{(p)} + \bar{\mE}_3^{(p)T}) + \sum_{p, q=2}^{\infty} \bar{\mE}_3^{(p)T} \mE_3^{(q)}~, \cr
Y &=& \eta + \sum_{p=2}^{\infty} Y^{(p)} = \eta + \sum_{p=2}^{\infty} (\bar{\mE}_1^{(p)} + \bar{\mE}_3^{(p)T}) + \sum_{p, q=2}^{\infty} \bar{\mE}_3^{(p)T} \mE_1^{(q)}~, \cr
Z &=& \eta + \sum_{p=2}^{\infty} Z^{(p)} = \eta + \sum_{p=2}^{\infty} (\bar{\mE}_1^{(p)} + \bar{\mE}_1^{(p)T}) + \sum_{p, q=2}^{\infty} \bar{\mE}_1^{(p)T} \mE_1^{(q)}~,
\eea
where $\mE_{q}^{(p)} (x, y)$, $X^{(p)} (x, y)$, $Y^{(p)} (x, y)$ and $Z^{(p)} (x, y)$ are contributions to the relevant matrices at the $p$-th order in $y$. The results up to $p=10$ are given below. The following notation will be used in writing down results for $X$, $Y$ and $Z$. Given a matrix of the form: $\bar R_m R_n \cdots = \eta R_{m} R_{n} \cdots$, we define,
\bea
\{\bar R_m R_n \cdots \} = \eta R_{m} R_{n} \cdots + \eta \cdots R_{n} R_{m}  ~.
\eea

\vspace{.2in}
\noindent
\underline{\bf Results for $\mE_1$:}
\bea
\mE^{(2)}_{1} &=& {1\over 6} \hR_2~, \cr
\mE^{(3)}_{1} &=& {1\over 12} \hR_3~, \cr
\mE^{(4)}_{1} &=& {1\over 40} \hR_4 + {1\over 120} \hR_2^2 ~, \cr
\mE^{(5)}_{1} &=& {1\over 180} \hR_5 + {1\over 180} \hR_3 \hR_2 +
{1\over 360} \hR_2 \hR_3 ~, \cr
\mE^{(6)}_{1} &=& {1\over 1008} \hR_6 + {1\over 504} \hR_4 \hR_2 +
{1\over 1680} \hR_2 \hR_4 + {1\over 504} \hR_3^2 + {1\over 5040} \hR_2^3~, \cr
\mE^{(7)}_{1} &=& {1\over 6720} \hR_7 + {1\over 2016} \hR_5 \hR_2
+ {1\over 1344} \hR_4 \hR_3 + {1\over 2240} \hR_3 \hR_4 + {1\over 6720} \hR_3 \hR_2^2
+ {1\over 10080} \hR_2 \hR_5 \cr
&& + {1\over 10080} \hR_2 \hR_3 \hR_2 + {1\over 20160} \hR_2^2 \hR_3 ~, \cr
\mE^{(8)}_{1} &=& {1\over 51840} \hR_8 + {1\over 10368} \hR_6 \hR_2 + {1\over 5184} \hR_5 \hR_3 + {1\over 5760} \hR_4^2 + {1\over 17280} \hR_4 \hR_2^2 \cr
&& + {1\over 12960} \hR_3 \hR_5 + {1\over 12960} \hR_3^2 \hR_2 + {1\over 25920} \hR_3 \hR_2 \hR_3 +
{1\over 72576} \hR_2 \hR_6 \cr
&& + {1\over 36288} \hR_2 \hR_4 \hR_2
+ {1\over 36288} \hR_2 \hR_3^2 + {1\over 120960} \hR_2^2 \hR_4 + {1\over 362880} \hR_2^4~,  \cr
\mE^{(9)}_{1} &=& {1\over 453600} \hR_9 + {1\over 64800} \hR_7 \hR_2 + {1\over 25920} \hR_6 \hR_3 + {1\over 21600} \hR_5 \hR_4 + {1\over 64800} \hR_5 \hR_2^2 + {1\over 32400} \hR_4 \hR_5 \cr
&& + {1\over 32400} \hR_4 \hR_3 \hR_2 + {1\over 64800} \hR_4 \hR_2 \hR_3 + {1\over 90720} \hR_3 \hR_6
+ {1\over 45360} \hR_3 \hR_4 \hR_2 + {1\over 45360} \hR_3^3 \cr
&& + {1\over 151200} \hR_3 \hR_2 \hR_4 + {1\over 453600}\hR_3 \hR_2^3 + {1\over 604800} \hR_2 \hR_7 + {1\over 181440} \hR_2 \hR_5 \hR_2 + {1\over 120960} \hR_2 \hR_4 \hR_3 \cr
&& + {1\over 201600} \hR_2 \hR_3 \hR_4 + {1\over 604800} \hR_2 \hR_3 \hR_2^2 + {1\over 907200} \hR_2^2 \hR_5
+ {1\over 907200} \hR_2^2 \hR_3 \hR_2 \cr
&& + {1\over 1814400} \hR_2^3 \hR_3~, \cr
\mE^{(10)}_{1} &=& {1\over 4435200} \hR_{10} + {1\over 475200} \hR_8 \hR_2 + {1\over 158400} \hR_7 \hR_3 + {1\over 105600} \hR_6 \hR_4 + {1\over 316800} \hR_6 \hR_2^2 \cr
&& + {1\over 118800} \hR_5^2 + {1\over 118800} \hR_5 \hR_3 \hR_2 + {1\over 237600} \hR_5 \hR_2 \hR_3
+ {1\over 221760} \hR_4 \hR_6 \cr
&& + {1\over 110880} \hR_4^2 \hR_2 + {1\over 110880} \hR_4 \hR_3^2 + {1\over 369600} \hR_4 \hR_2 \hR_4
+ {1\over 1108800} \hR_4 \hR_2^3
+ {1\over 739200} \hR_3 \hR_7 \cr
&& + {1\over 221760} \hR_3 \hR_5 \hR_2 + {1\over 147840} \hR_3 \hR_4 \hR_3 + {1\over 246400} \hR_3^2 \hR_4 + {1\over 739200} \hR_3^2 \hR_2^2 + {1\over 1108800} \hR_3 \hR_2 \hR_5 \cr
&& + {1\over 1108800} \hR_3 \hR_2 \hR_3 \hR_2 + {1\over 2217600} \hR_3 \hR_2^2 \hR_3
+ {1\over 5702400} \hR_2 \hR_8 \cr
&& + {1\over 1140480} \hR_2 \hR_6 \hR_2 + {1\over 570240} \hR_2 \hR_5 \hR_3 + {1\over 633600} \hR_2 \hR_4^2 \cr
&& + {1\over 1900800} \hR_2 \hR_4 \hR_2^2 + {1\over 1425600} \hR_2 \hR_3 \hR_5
+ {1\over 1425600} \hR_2 \hR_3^2 \hR_2 \cr
&& + {1\over 2851200} \hR_2 \hR_3 \hR_2 \hR_3 + {1\over 7983360} \hR_2^2 \hR_6
+ {1\over 3991680} \hR_2^2 \hR_4 \hR_2 \cr
&& + {1\over 3991680} \hR_2^2 \hR_3^2
+ {1\over 13305600} \hR_2^3 \hR_4 + {1\over 39916800} \hR_2^5 ~.
\nonumber
\eea

\vspace{.2in}
\noindent
\underline{\bf Results for $\mE_3$:}
\bea
\mE^{(2)}_{3} &=& {1\over 2} \hR_2~, \cr
\mE^{(3)}_{3} &=& {1\over 6} \hR_3~, \cr
\mE^{(4)}_{3} &=& {1\over 24} \hR_4 + {1\over 24} \hR_2^2 ~, \cr
\mE^{(5)}_{3} &=& {1\over 120} \hR_5 + {1\over 40} \hR_3 \hR_2 +
{1\over 120} \hR_2 \hR_3 ~, \cr
\mE^{(6)}_{3} &=& {1\over 720} \hR_6 + {1\over 120} \hR_4 \hR_2 +
{1\over 720} \hR_2 \hR_4 + {1\over 180} \hR_3^2 + {1\over 720} \hR_2^3~, \cr
\mE^{(7)}_{3} &=& {1\over 5040} \hR_7 + {1\over 504} \hR_5 \hR_2
+ {1\over 504} \hR_4 \hR_3 + {1\over 1008} \hR_3 \hR_4 + {1\over 1008} \hR_3 \hR_2^2 \cr
&& + {1\over 5040} \hR_2 \hR_5 + {1\over 1680} \hR_2 \hR_3 \hR_2 + {1\over 5040} \hR_2^2 \hR_3 ~, \cr
\mE^{(8)}_{3} &=& {1\over 40320} \hR_8 + {1\over 2688} \hR_6 \hR_2 + {1\over 2016} \hR_5 \hR_3 + {1\over 2688} \hR_4^2 + {1\over 2688} \hR_4 \hR_2^2 \cr
&& + {1\over 6720} \hR_3 \hR_5 + {1\over 2240} \hR_3^2 \hR_2
+ {1\over 6720} \hR_3 \hR_2 \hR_3 + {1\over 40320} \hR_2 \hR_6 \cr
&& + {1\over 6720} \hR_2 \hR_4 \hR_2 + {1\over 10080} \hR_2 \hR_3^2 + {1\over 40320} \hR_2^2 \hR_4 + {1\over 40320} \hR_2^4~,  \cr
\mE^{(9)}_{3} &=& {1\over 362880} \hR_9 + {1\over 17280} \hR_7 \hR_2 + {1\over 10368} \hR_6 \hR_3 + {1\over 10368} \hR_5 \hR_4 + {1\over 10368} \hR_5 \hR_2^2 \cr
&& + {1\over 17280} \hR_4 \hR_5 + {1\over 5760} \hR_4 \hR_3 \hR_2
+ {1\over 17280} \hR_4 \hR_2 \hR_3
+ {1\over 51840} \hR_3 \hR_6 \cr
&& + {1\over 8640} \hR_3 \hR_4 \hR_2 + {1\over 12960} \hR_3^3
+ {1\over 51840} \hR_3 \hR_2 \hR_4 + {1\over 51840} \hR_3 \hR_2^3 \cr
&& + {1\over 362880} \hR_2 \hR_7 + {1\over 36288} \hR_2 \hR_5 \hR_2
+ {1\over 36288} \hR_2 \hR_4 \hR_3 + {1\over 72576} \hR_2 \hR_3 \hR_4 \cr
&& + {1\over 72576} \hR_2 \hR_3 \hR_2^2 + {1\over 362880} \hR_2^2 \hR_5
+ {1\over 120960} \hR_2^2 \hR_3 \hR_2 \cr
&& + {1\over 362880} \hR_2^3 \hR_3~, \cr
\mE^{(10)}_{3} &=& {1\over 3628800} \hR_{10} + {1\over 129600} \hR_8 \hR_2 + {1\over 64800} \hR_7 \hR_3 + {1\over 51840} \hR_6 \hR_4 + {1\over 51840} \hR_6 \hR_2^2 \cr
&& + {1\over 64800} \hR_5^2 + {1\over 21600} \hR_5 \hR_3 \hR_2 + {1\over 64800} \hR_5 \hR_2 \hR_3 + {1\over 129600} \hR_4 \hR_6 \cr
&& + {1\over 21600} \hR_4^2 \hR_2 + {1\over 32400} \hR_4 \hR_3^2
+ {1\over 129600} \hR_4 \hR_2 \hR_4 \cr
&& + {1\over 129600} \hR_4 \hR_2^3 + {1\over 453600} \hR_3 \hR_7 + {1\over 45360} \hR_3 \hR_5 \hR_2 \cr
&& + {1\over 45360} \hR_3 \hR_4 \hR_3 + {1\over 90720} \hR_3^2 \hR_4 + {1\over 90720} \hR_3^2 \hR_2^2 \cr
&& + {1\over 453600} \hR_3 \hR_2 \hR_5 + {1\over 151200} \hR_3 \hR_2 \hR_3 \hR_2
+ {1\over 453600} \hR_3 \hR_2^2 \hR_3 \cr
&& + {1\over 3628800} \hR_2 \hR_8 + {1\over 241920} \hR_2 \hR_6 \hR_2
+ {1\over 181440} \hR_2 \hR_5 \hR_3 \cr
&& + {1\over 241920} \hR_2 \hR_4^2 + {1\over 241920} \hR_2 \hR_4 \hR_2^2
+ {1\over 604800} \hR_2 \hR_3 \hR_5 \cr
&& + {1\over 201600} \hR_2 \hR_3^2 \hR_2 + {1\over 604800} \hR_2 \hR_3 \hR_2 \hR_3
+ {1\over 3628800} \hR_2^2 \hR_6 \cr
&& + {1\over 604800} \hR_2^2 \hR_4 \hR_2 + {1\over 907200} \hR_2^2 \hR_3^2
+ {1\over 3628800} \hR_2^3 \hR_4 \cr
&& + {1\over 3628800} \hR_2^5 ~.
\nonumber
\eea

\vspace{.2in}
\noindent
\underline{\bf Results for $G^{\parallel}$:}
\bea
X^{(2)} &=& \bar{\hR}_2 ~, \cr
X^{(3)} &=& {1\over 3} \bar{\hR}_3 ~, \cr
X^{(4)} &=& {1\over 12} \bar{\hR}_4 + {1\over 3}\bar{\hR}_2 \hR_2~, \cr
X^{(5)} &=& {1\over 60} \bar{\hR}_5 + {7\over 60} \{\bar{\hR}_3 \hR_2\}~, \cr
X^{(6)} &=& {1\over 360} \bar{\hR}_6 + {11\over 360} \{\bar{\hR}_4 \hR_2\}
+ {7\over 180} \bar{\hR}_3 \hR_3 + {2\over 45} \bar{\hR}_2 \hR_2^2~, \cr
X^{(7)} &=& {1\over 2520} \bar{\hR}_7 + {2\over 315}\{\bar{\hR}_5 \hR_2\} + {5\over 504} \{\bar{\hR}_4 \hR_3\} + {31\over 2520}\{\bar{\hR}_3 \hR_2^2\} + {11\over 420} \bar{\hR}_2 \hR_3 \hR_2 ~, \cr
X^{(8)} &=& {1\over 20160} \bar{\hR}_8 + {11\over 10080} \{\bar{\hR}_6 \hR_2\}
+ {41 \over 20160} \{\bar{\hR}_5 \hR_3\} + {5 \over 2016} \bar{\hR}_4 \hR_4
+ {19 \over 6720} \{\bar{\hR}_4 \hR_2^2\} \cr
&& + {41 \over 20160} \bar{\hR}_3 \hR_5 + {151 \over 20160}\{\bar{\hR}_3 \hR_3 \hR_2 \}
+ {31 \over 10080} \bar{\hR}_3 \hR_2 \hR_3 + {29 \over 3360} \bar{\hR}_2 \hR_4 \hR_2
+ {1\over 315} \bar{\hR}_2 \hR_2^3~,\cr
X^{(9)} &=& {1 \over 181440} \bar{\hR}_9 + {29 \over 181440} \{\bar{\hR}_7 \hR_2 \} + {1\over 2880} \{\bar{\hR}_6 \hR_3 \} + {13 \over 25920} \{\bar{\hR}_5 \hR_4 \} + {11 \over 20160} \{\bar{\hR}_5 \hR_2^2 \} \cr
&&
+ {313 \over 181440} \{\bar{\hR}_4 \hR_3 \hR_2 \}
+ {17 \over 25920} \{\bar{\hR}_4 \hR_2 \hR_3 \}
+ {229 \over 90720} \{\bar{\hR}_3 \hR_4 \hR_2 \}
+ {13 \over 6480} \bar{\hR}_3\hR_3^2 \cr
&&
+ {127 \over 181440} \{\bar{\hR}_3\hR_2^3\}
+ {37 \over 18144} \bar{\hR}_2\hR_5\hR_2
+ {337 \over 181440} \{\bar{\hR}_2 \hR_3 \hR_2^2 \} \cr
X^{(10)} &=& {1\over 1814400} \bar{\hR}_{10}
+ {37 \over 1814400} \{\bar{\hR}_8\hR_2 \}
+ {23 \over 453600} \{\bar{\hR}_7 \hR_3 \}
+ {11 \over 129600} \{\bar{\hR}_6 \hR_4\} \cr
&& + {163\over 1814400}\{\bar{\hR}_6\hR_2^2 \}
+ {13\over 129600} \bar{\hR}_5 \hR_5
+ {1\over 3024} \{ \bar{\hR}_5\hR_3\hR_2 \}
+ {109\over 907200} \{ \bar{\hR}_5\hR_2\hR_3 \} \cr
&&+ {353\over 604800} \{ \bar{\hR}_4\hR_4\hR_2 \}
+ {199\over 453600} \{\bar{\hR}_4 \hR_3^2 \}
+ {17\over 129600} \bar{\hR}_4\hR_2\hR_4
+ {247\over 1814400} \{ \bar{\hR}_4\hR_2^3 \} \cr
&& + {11\over 18144} \{\bar{\hR}_3\hR_5\hR_2\} + {2\over 2835} \bar{\hR}_3\hR_4\hR_3
+ {13\over 28350} \{\bar{\hR}_3\hR_3\hR_2^2 \}
+ {59\over 151200}\bar{\hR}_3\hR_2\hR_3\hR_2 \cr
&&+ {127\over 907200} \bar{\hR}_3\hR_2^2\hR_3 + {23\over 60480} \bar{\hR}_2\hR_6\hR_2
+ {53\over 86400} \{\bar{\hR}_2\hR_4\hR_2^2 \} + {109\over 100800} \bar{\hR}_2\hR_3^2\hR_2 \cr
&& + {59\over 151200} \bar{\hR}_2\hR_3\hR_2\hR_3 + {2\over 14175} \bar{\hR}_2\hR_2^4 ~.
\nonumber
\eea

\vspace{.2in}
\noindent
\underline{\bf Results for $H$:}
\bea
Y^{(2)} &=& {2\over 3} \bar{\hR}_2~, \cr
Y^{(3)} &=& {1\over 4} \bar{\hR}_3~, \cr
Y^{(4)} &=& {1\over 15} \bar{\hR}_4 + {2\over 15} \bar{\hR}_2 \hR_2~, \cr
Y^{(5)} &=& {1\over 72}\bar{\hR}_5 + {1\over 24} \bar{\hR}_3 \hR_2 + {5\over 72} \bar{\hR}_2 \hR_3~, \cr
Y^{(6)} &=& {1\over 420} \bar{\hR}_6 + {1\over 63} \bar{\hR}_4 \hR_2 + {3\over 140} \bar{\hR}_3 \hR_3 +  {3\over 140} \bar{\hR}_2 \hR_4 + {5\over 504} \bar{\hR}_2 \hR_2^2 ~, \cr
Y^{(7)} &=& {1\over 2880} \bar{\hR}_7
+ {1\over 480} \bar{\hR}_5\hR_2
+ {1\over 192} \bar{\hR}_4 \hR_3
+ {19\over 2880} \bar{\hR}_3 \hR_4
+ {1\over 320} \bar{\hR}_3 \hR_2^2
+ {7\over 1440} \bar{\hR}_2 \hR_5 \cr
&& + {11\over 1440} \bar{\hR}_2 \hR_3 \hR_2
+ {17\over 2880} \bar{\hR}_2 \hR_2 \hR_3 ~, \cr
Y^{(8)} &=& {1\over 22680} \bar{\hR}_8
+ {1\over 2835} \bar{\hR}_6\hR_2
+ {47\over 45360} \bar{\hR}_5\hR_3
+ {1\over 630} \bar{\hR}_4\hR_4
+ {1\over 1512} \bar{\hR}_4\hR_2^2 \cr
&& + {17\over 11340} \bar{\hR}_3\hR_5
+ {23\over 11340} \bar{\hR}_3\hR_3\hR_2
+ {61\over 45360} \bar{\hR}_3\hR_2\hR_3
+ {1\over 1134}\bar{\hR}_2\hR_6
+ {29\over 11340} \bar{\hR}_2\hR_4\hR_2 \cr
&& +  {23\over 6480} \bar{\hR}_2\hR_3^2
+ {13\over 7560} \bar{\hR}_2\hR_2\hR_4
+ {2\over 2835} \bar{\hR}_2\hR_2^3~, \cr
Y^{(9)} &=& {1\over 201600} \bar{\hR}_9
+ {31\over 604800} \bar{\hR}_7\hR_2
+ {1\over 5760} \bar{\hR}_6\hR_3
+ {1\over 3200} \bar{\hR}_5\hR_4
+ {73\over 604800} \bar{\hR}_5\hR_2^2 \cr
&&
+ {31\over 86400} \bar{\hR}_4\hR_5
+ {89\over 201600} \bar{\hR}_4\hR_3\hR_2
+ {23\over 86400} \bar{\hR}_4\hR_2\hR_3
+ {11\over 40320} \bar{\hR}_3\hR_6
+ {43\over 60480} \bar{\hR}_3\hR_4\hR_2 \cr
&&
+ {1\over 1120} \bar{\hR}_3\hR_3^2
+ {1\over 2688} \bar{\hR}_3\hR_2\hR_4
+ {17\over 120960} \bar{\hR}_3\hR_2^3
+  {3\over 22400} \bar{\hR}_2\hR_7
+ {37\over 60480} \bar{\hR}_2\hR_5\hR_2 \cr
&&
+ {1\over 840} \bar{\hR}_2\hR_4\hR_3
+ {23\over 22400} \bar{\hR}_2\hR_3\hR_4
+ {79\over 201600} \bar{\hR}_2\hR_3\hR_2^2
+ {229\over 604800} \bar{\hR}_2\hR_2\hR_5 \cr
&&
+ {31\over 67200} \bar{\hR}_2\hR_2\hR_3\hR_2
+ {167\over 604800} \bar{\hR}_2\hR_2^2\hR_3~, \cr
Y^{(10)} &=& {1\over 1995840} \bar{\hR}_{10}
+ {13\over 1995840} \bar{\hR}_8\hR_2
+ {5\over 199584} \bar{\hR}_7\hR_3
+ {37\over 712800} \bar{\hR}_6\hR_4
+ {191\over 9979200} \bar{\hR}_6\hR_2^2 \cr
&&+ {1\over 14256} \bar{\hR}_5\hR_5
+ {1\over 12320} \bar{\hR}_5\hR_3\hR_2
+ {23\over 498960} \bar{\hR}_5\hR_2\hR_3
+ {13\over 199584} \bar{\hR}_4\hR_6
+ {1\over 6336} \bar{\hR}_4\hR_4\hR_2\cr
&&
+ {37\over 199584} \bar{\hR}_4\hR_3^2
+ {349\over 4989600} \bar{\hR}_4\hR_2\hR_4
+ {251\over 9979200} \bar{\hR}_4\hR_2^3
+ {83\over 1995840} \bar{\hR}_3\hR_7 \cr
&&
+ {5\over 28512} \bar{\hR}_3\hR_5\hR_2
+ {127\over 399168} \bar{\hR}_3\hR_4\hR_3
+ {2477\over 9979200} \bar{\hR}_3\hR_3\hR_4
+ {899\over 9979200} \bar{\hR}_3\hR_3\hR_2^2 \cr
&&
+ {79\over 997920} \bar{\hR}_3\hR_2\hR_5
+ {1\over 11088} \bar{\hR}_3\hR_2\hR_3\hR_2
+ {101\over 1995840} \bar{\hR}_3\hR_2^2\hR_3
+ {1\over 57024} \bar{\hR}_2\hR_8 \cr
&&
+ {23\over 199584} \bar{\hR}_2\hR_6\hR_2
+ {19\over 66528} \bar{\hR}_2\hR_5\hR_3
+ {163\over 475200} \bar{\hR}_2 \hR_4^2
+ {139\over 1108800} \bar{\hR}_2\hR_4\hR_2^2 \cr
&&
+ {1\over 4455} \bar{\hR}_2\hR_3\hR_5
+ {257\over 997920} \bar{\hR}_2\hR_3^2\hR_2
+ {29\over 199584} \bar{\hR}_2\hR_3\hR_2\hR_3
+ {1\over 14784} \bar{\hR}_2\hR_2\hR_6 \cr
&&
+ {65\over 399168} \bar{\hR}_2\hR_2\hR_4\hR_2
+ {19\over 99792} \bar{\hR}_2\hR_2\hR_3^2
+ {713\over 9979200} \bar{\hR}_2 \hR_2^2\hR_4
+ {4\over 155925} \bar{\hR}_2 \hR_2^4 ~,
\nonumber
\eea

\vspace{.2in}
\noindent
\underline{\bf Results for $G^{\perp}$:}
\bea
Z^{(2)} &=& {1\over 3} \bar{\hR}_2 ~, \cr
Z^{(3)} &=& {1\over 6} \bar{\hR}_3 ~, \cr
Z^{(4)} &=& {1\over 20} \bar{\hR}_4 + {2\over 45} \bar{\hR}_2\hR_2 ~, \cr
Z^{(5)} &=& {1\over 90} \bar{\hR}_5 + {1\over 45}
\{\bar{\hR}_3\hR_2 \}~, \cr
Z^{(6)} &=& {1\over 504} \bar{\hR}_6 + {17\over 2520} \{\bar{\hR}_4\hR_2 \}
+ {11\over 1008} \bar{\hR}_3\hR_3 + {1\over 315} \bar{\hR}_2^3 ~, \cr
Z^{(7)} &=& {1\over 3360} \bar{\hR}_7 + {23\over 15120} \{\bar{\hR}_5 \hR_2 \}
+ {11\over 3360} \{\bar{\hR}_4\hR_3 \} + {41\over 30240} \{\bar{\hR}_3\hR_2^2 \}
+ {31\over 15120} \bar{\hR}_2\hR_3\hR_2 \cr
Z^{(8)} &=&  {1\over 25920} \bar{\hR}_8 + {5\over 18144} \{ \bar{\hR}_6\hR_2 \}
+ {19\over 25920} \{\bar{\hR}_5\hR_3 \} + {7\over 7200} \bar{\hR}_4\hR_4
+ {113\over 302400} \{\bar{\hR}_4\hR_2^2\} \cr
&&
+ {163\over 181440} \{\bar{\hR}_3\hR_3\hR_2 \}
+ {7\over 12960} \bar{\hR}_3\hR_2\hR_3 + {13\over 18144} \bar{\hR}_2\hR_4\hR_2
+ {2\over 14175} \bar{\hR}_2\hR_2^3 ~, \cr
Z^{(9)} &=&  {1\over 226800} \bar{\hR}_9 + {19\over 453600} \{ \bar{\hR}_7 \hR_2 \}
+ {1\over 7560} \{\bar{\hR}_6\hR_3 \} + {7\over 32400} \{\bar{\hR}_5\hR_4 \} \cr
&&
+ {1\over 12600} \{\bar{\hR}_5\hR_2^2 \} + {113\over 453600} \bar{\hR}_4\hR_3\hR_2
+ {2\over 14175} \bar{\hR}_4\hR_2\hR_3 + {29\over 90720} \bar{\hR}_3\hR_4\hR_2 \cr
&&
+ {17\over 45360} \bar{\hR}_3\hR_3^2 + {2\over 14175} \bar{\hR}_3\hR_2\hR_4
+  {23\over 453600} \{\bar{\hR}_3\hR_2^3 \} + {1\over 5670} \bar{\hR}_2\hR_5\hR_2 \cr
&&
+ {29\over 90720} \bar{\hR}_2\hR_4\hR_3 + {113\over 453600} \bar{\hR}_2\hR_3\hR_4
+ {41\over 453600} \{\bar{\hR}_2\hR_3\hR_2^2 \}~, \cr
Z^{(10)} &=& {1\over 2217600} \bar{\hR}_{10} + {47\over 8553600} \{\bar{\hR}_8\hR_2 \}
+{89 \over 4435200} \{ \bar{\hR}_7 \hR_3 \}
+  {43\over 1108800} \{\bar{\hR}_6\hR_4 \} + {829\over 59875200} \{\bar{\hR}_6\hR_2^2 \} \cr
&&
+ {17\over 356400} \bar{\hR}_5\hR_5 + {113\over 2138400} \bar{\hR}_5\hR_3\hR_2
+ {23\over 798336} \bar{\hR}_5\hR_2\hR_3 + {593\over 6652800} \{\bar{\hR}_4 \hR_4\hR_2 \} \cr
&&
+ {443\over 4435200} \{ \bar{\hR}_4\hR_3^2 \} + {13\over 369600} \bar{\hR}_4\hR_2\hR_4
+ {7\over 570240} \{ \bar{\hR}_4\hR_2^3 \}  \cr
&&
+ {191\over 2395008} \bar{\hR}_3\hR_5\hR_2 + {61\over 443520} \bar{\hR}_3\hR_4\hR_3
+ {601\over 17107200} \{\bar{\hR}_3\hR_3\hR_2^2 \} + {23\over 798336} \bar{\hR}_3\hR_2\hR_5 \cr
&&
+ {1879\over 59875200} \bar{\hR}_3\hR_2\hR_3\hR_2 + {337\over 19958400} \bar{\hR}_3\hR_2^2\hR_3 + {29\over 855360} \bar{\hR}_2\hR_6\hR_2 + {191\over 2395008} \bar{\hR}_2\hR_5\hR_3 \cr
&&
+ {1889\over 59875200} \bar{\hR}_2\hR_4\hR_2^2  + {113\over 2138400} \bar{\hR}_2\hR_3\hR_5
+ {31\over 534600} \bar{\hR}_2\hR_3^2\hR_2 + {1879\over 59875200} \bar{\hR}_2\hR_3\hR_2\hR_3 \cr
&&
+ {1889\over 59875200} \bar{\hR}_2\hR_2\hR_4\hR_2 + {2\over 467775} \bar{\hR}_2\hR_2^4 ~.
\nonumber
\eea


\begin{thebibliography}{99}

\bibitem{bishop}
R.~L.~Bishop and R.~J.~Crittenden, ``Geometry of Manifolds," New York: Academic Press, 1964.

\bibitem{eisenhart}
L. ~Eisenhart, ``Riemannian Geometry," Princeton Univ. Press, Princeton, N.J., 1965.

\bibitem{hawking} 
S.~W.~Hawking and G.~F.~R.~Ellis,
``The Large scale structure of space-time,'' Cambridge University Press, Cambridge, 1973.

\bibitem{wald}
R. ~M. ~Wald, ``General Relativity," Chicago, Usa: Univ. Pr. ( 1984) 491p.
  
\bibitem{QFT1}
T.~S.~Bunch and L.~Parker,
``Feynman Propagator In Curved Space-Time: A Momentum Space Representation,''
Phys.\ Rev.\  D {\bf 20}, 2499 (1979).

\bibitem{QFT2}
M. L\"{u}scher, 
``Dimensional regularisation in the presence of large background fields,"
Ann. \ Phys. \ (N.Y.) {\bf 142} (1982) 359.

\bibitem{sigma1}
D.~Friedan,
``Nonlinear Models In Two Epsilon Dimensions,''
Phys.\ Rev.\ Lett.\  {\bf 45}, 1057 (1980).

\bibitem{sigma2}
D.~H.~Friedan,
``Nonlinear Models In Two + Epsilon Dimensions,''
Annals Phys.\  {\bf 163}, 318 (1985).

\bibitem{sigma3}
L.~Alvarez-Gaume, D.~Z.~Freedman and S.~Mukhi,
``The Background Field Method And The Ultraviolet Structure Of The
Supersymmetric Nonlinear Sigma Model,''
Annals Phys.\  {\bf 134}, 85 (1981).

\bibitem{sigma4}
B.~E.~Fridling and A.~E.~M.~van de Ven,
``Renormalization Of Generalized Two-Dimensional Nonlinear Sigma Models,''
Nucl.\ Phys.\  B {\bf 268}, 719 (1986).


\bibitem{fermi}
E. Fermi, ``On Phenomena Occurring Close to a World Line," Rend. \ Lincei, {\bf 31} 17-24  (1922). 

\bibitem{fermi-gen1}  
L. ~O'Raifeartaigh, ``Fermi Coordinates," Proc. \ Roy. \ Irish \ Acad. {\bf A59}, 2 (1958).

\bibitem{fermi-gen2}  
J.~L.~Synge,
``Relativity: The General Theory,''
{\it  North-Holland, Amsterdam, 1960}.

\bibitem{fermi-gen3}  
F. K. Manasse and C. W. Misner, 
``Fermi Normal Coordinates and Some Basic Concepts in Differential Geometry,"
J. \ Math. \ Phys. {\bf 4}, 735 (1963).

\bibitem{FS}
P.~S. Florides and J.~L.~Synge,
``Coordinate Conditions in a Riemannian Space for Coordinates Based on a Subspace,''
Proceedings of the Royal Society of London. Series A, Mathematical and Physical Sciences
Vol. 323, No. 1552 (May 25, 1971), pp. 1-10
Published by: The Royal Society
Stable URL: http://www.jstor.org/stable/77913

\bibitem{spivak}
M.~Spivak, ``A Comprehensive Introduction to Differential Geometry," vol. I, second edition, Publish or Perish, Inc., Wilmington, 1979.

\bibitem{bredon}
G. E. Bredon, Topology and Geometry, Graduate Texts in Math. 139, Springer, New York, 1993.

\bibitem{tubes}
A.~Gray, ``Tubes," Second Edition (Progress in Mathematics). Addison-Wesley Publishing Company, New York, 1990.

A review by J.~D.~Moore can be found here:\\
http://www.ams.org/journals/bull/1992-27-02/S0273-0979-1992-00312-9/home.html

\bibitem{tube-vol1}
H.~Hotelling,  ``Tubes and spheres in n-spaces, and a class of statistical problems", American Journal of Mathematics, {\bf 61} 440-460 (1939).

\bibitem{tube-vol2}
H.~ Weyl, ``On the volume of tubes," American Journal of Mathematics, {\bf 61} 461-472 (1939).

\bibitem{loader}
C.~Loader, ``The Volume-of-Tubes Formula: Computational Methods and Statistical Applications,"
[arXiv:math/0511502v1 [math.ST]].

\bibitem{li1}
Li, W. Q. and Ni, W. T., ``Coupled inertial and gravitational effects in the proper reference frame of an accelerated, rotating observer," J. \ Math. \ Phys. {\bf 20} 1473-80 (1979).

\bibitem{li2}
Li, W. Q. and Ni, W. T., ``Expansions of the affinity, metric and geodesic equations inFermi normal coordinates about a geodesic," J. \ Math. \ Phys. {\bf 20} 1925-9 (1979).

\bibitem{bini1}
Bini, D., Geralico, A., Jantzen, R., ``Kerr metric, static observers and Fermi coordinates," Class. \ Quant. \ Grav. {\bf 22}, 4729-4742 (2005).

\bibitem{bini2}
Bini, D., Geralico, A., and Jantzen, R., ``Fermi coordinates in Schwarzschild spacetime: closed form expressions," Gen. \ Rel. and Grav. {\bf 43},  1837-1853, (2011).

\bibitem{klein1}
Klein, D., Collas, P. ``General transformation formulas for Fermi-Walker coordinates," Class. \ Quantum \ Gravit. {\bf 25}, 145019 (2008).

\bibitem{klein2}
Klein, D., Randles, E. ``Fermi coordinates, simultaneity, and expanding space in Robertson-Walker cosmologies," Ann. \ Henri \ Poincar {\bf 12}, 303-328 (2011).

\bibitem{FNC-GR1}
K.~P.~Marzlin,
``Fermi coordinates for weak gravitational fields,''
Phys.\ Rev.\  D {\bf 50}, 888 (1994)
[arXiv:gr-qc/9403044];

\bibitem{FNC-GR2}
M.~Ishii, M.~Shibata and Y.~Mino,
``Black hole tidal problem in the Fermi normal coordinates,''
Phys.\ Rev.\  D {\bf 71}, 044017 (2005)
[arXiv:gr-qc/0501084];

\bibitem{FNC-GR3}
C.~Chicone and B.~Mashhoon,
``Explicit Fermi coordinates and tidal dynamics in de Sitter and Goedel spacetimes,''
Phys.\ Rev.\  D {\bf 74}, 064019 (2006)
[arXiv:gr-qc/0511129].

\bibitem{living-rev}
E.~Poisson, A.~Pound and I.~Vega,
``The Motion of point particles in curved spacetime,''
Living Rev.\ Rel.\  {\bf 14}, 7 (2011)
[arXiv:1102.0529 [gr-qc]].

\bibitem{parker1} 
L.~Parker,
``One-Electron Atom in Curved Space-Time,"
Phys. \ Rev. \ Lett. {\bf 44}, 1559-1562 (1980) .

\bibitem{parker2}
L.~Parker,
``One Electron Atom As A Probe Of Space-time Curvature,''
Phys.\ Rev.\ D {\bf 22}, 1922 (1980).

\bibitem{parker3}
E.~Fischbach, B.~S.~Freeman and W.~-K.~Cheng,
``General Relativistic Effects In Hydrogenic Systems,''
Phys.\ Rev.\ D {\bf 23}, 2157 (1981)
[Erratum-ibid.\ D {\bf 24}, 1702 (1981)].

\bibitem{parker4}
L.~Parker and L.~O.~Pimentel,
``Gravitational Perturbation Of The Hydrogen Spectrum,''
Phys.\ Rev.\ D {\bf 25}, 3180 (1982).

\bibitem{gill}
E. ~Gill, G. ~Wunner, M. ~Soffel and H. ~Ruder, 
``On hydrogen-like atoms in strong gravitational fields,"
Class. \ Quantum \ Grav. {\bf 4}, (1987) 1031.

\bibitem{pinto}
F. ~Pinto, 
``Rydberg atoms in curved space-time,"
Phys. \ Rev. \ Lett. {\bf 70} (1993) 3839.

\bibitem{audretsch}
J.~Audretsch and K.~P.~Marzlin,
``Ramsey fringes in atomic interferometry: Measurability of the influence of space-time curvature,''
Phys.\ Rev.\ A {\bf 50}, 2080 (1994)
[gr-qc/9310029].

\bibitem{zhao}
Z.~-H.~Zhao, Y.~-X.~Liu and X.~-G.~Li,
``The Energy-Level Shifts of a Stationary Hydrogen Atom in Static External Gravitational Field with Schwarzschild Geometry,''
Phys.\ Rev.\ D {\bf 76}, 064016 (2007)
[arXiv:0705.1571 [gr-qc]].

\bibitem{moradi}
S.~Moradi and E.~Aboualizadeh,
``Hydrogen atom and its energy level shifts in de Sitter universe,''
Gen.\ Rel.\ Grav.\  {\bf 42}, 435 (2010).

\bibitem{caicedo}
J.~A.~Caicedo and L.~F.~Urrutia,
``Relativistic Two-Body Coulomb-Breit Hamiltonian in an External Weak Gravitational Field,''
Phys.\ Lett.\ B {\bf 705}, 143 (2011)
[arXiv:1110.0109 [gr-qc]].

\bibitem{collas} 
P.~Collas and D.~Klein,
``A Statistical mechanical problem in Schwarzschild spacetime,''
Gen.\ Rel.\ Grav.\  {\bf 39}, 737 (2007)
[gr-qc/0603086].

\bibitem{relativistic-bound1} 
N.~Nakanishi,
``A General survey of the theory of the Bethe-Salpeter equation,''
Prog.\ Theor.\ Phys.\ Suppl.\  {\bf 43}, 1 (1969);

\bibitem{relativistic-bound2}
H.~W.~Crater and J.~Schiermeyer, 
``Applications of two-body Dirac equations to the meson spectrum with three versus two covariant interactions, SU(3) mixing, and comparison to a quasipotential approach,''
Phys. \ Rev. {D 82} 094020 (2010).

\bibitem{semi-classical}
P.~Mukhopadhyay,
``On a semi-classical limit of loop space quantum mechanics,''
ISRN High Energy Phys.\  {\bf 2013}, 398030 (2013) [arXiv:1202.2735 [hep-th]].

\bibitem{dwv1} 
P.~Mukhopadhyay,
``On a coordinate independent description of string worldsheet theory,''
Annales Henri Poincaré (June 2013), DOI: 10.1007/s00023-013-0256-6,
arXiv:0912.3987 [hep-th];

\bibitem{dwv2}
P.~Mukhopadhyay,
``DeWitt-Virasoro construction,''
Pramana {\bf 76}, 407 (2011)
[arXiv:1010.0930 [hep-th]].

\bibitem{dwv3}
P.~Mukhopadhyay,
``DeWitt-Virasoro construction in tensor representations,''
Adv.\ High Energy Phys.\  {\bf 2012}, 415634 (2012)
[arXiv:1004.2396 [hep-th]].

\bibitem{dacosta}
R.~C.~T.~da Costa, ``Constraints in quantum mechanics," Phys. Rev. A {\bf 25}, 2893-2900 (1982) . 

\bibitem{maraner}
P.~Maraner, ``A complete perturbative expansion for quantum mechanics with constraints," J. Phys. A {\bf 63}, 2939-2951 (1995).

\bibitem{mitchell}
K.~A.~Mitchell, ``Gauge fields and extrapotentials in constrained quantum systems," Phys. Rev. A {\bf 63}, 042112 (2001). 

\bibitem{wachsmuth1}
J.~Wachsmuth, S.~Teufel, ``Effective Hamiltonians for Constrained Quantum Systems," arXiv:0907.0351v3 [math-ph]. 

\bibitem{wachsmuth2}
J.~Wachsmuth, S.~Teufel, ``Constrained quantum systems as an adiabatic problem," Phy. Rev. A {\bf 82}, 022112 (2010). 

\bibitem{muller}
U.~Muller, C.~Schubert, A.~M.~E.~van de Ven,
``A Closed formula for the Riemann normal coordinate expansion,''
Gen.\ Rel.\ Grav.\  {\bf 31}, 1759-1768 (1999).
[gr-qc/9712092].

\bibitem{schwinger}
J.~ Schwinger, 
``On Gauge Invariance and Vacuum Polarization,"
Phys. \ Rev. {\bf 82} (1951) 664.

\bibitem{shifman}
M.~A.~Shifman,
``Wilson Loop In Vacuum Fields,''
Nucl.\ Phys.\  B {\bf 173}, 13 (1980).

\bibitem{nakahara}
M.~Nakahara,
``Geometry, Topology and Physics,'' Second Edition (Graduate Student Series in Physics).

\bibitem{exactFermi}
D.~Klein and P.~Collas,
``Exact Fermi coordinates for a class of spacetimes,''
J.\ Math.\ Phys.\  {\bf 51}, 022501 (2010)
[arXiv:0912.2779 [math-ph]].

\bibitem{progress} 
P.~Mukhopadhyay,
``Tubular expansion in multi-particle configuration space and loop space,''
{\it work in progress}.





\end{thebibliography}
\end{document}